\documentclass[pre,reprint]{revtex4-1}
\usepackage{bm,epsfig,graphicx,color,amssymb,amsmath}

\begin{document}

\title{Geometric universality of plasmon modes in graphene nanoribbon arrays}

\author{Kirill A. Velizhanin}
\email{kirill@lanl.gov}
\affiliation{Theoretical Division, Los Alamos National Laboratory, Los Alamos, New Mexico 87545, USA}

\begin{abstract}
Graphene plasmonics is a rapidly growing field with multiple potential applications. One of the standard ways to study plasmons in graphene is by fabricating an array of graphene nanoribbons where nanoribbon edges provide the efficient photon-plasmon coupling. We systematically analyze the problem of optical plasmonic response in such systems and demonstrate the purely geometric nature of the size quantization condition for graphene plasmons.  Accurate numerical calculations allowed us to tabulate the universal geometric parameters of plasmon size quantization, which is expected to become useful in analysis of experimental data on plasmonic response of graphene nanoribbons. A simple analytical theory has also been developed which accurately reproduces all the qualitative features of optical plasmonic response of graphene nanoribbons.  
\end{abstract}

\maketitle

\section{Introduction}\label{sec:int}

The study of infrared plasmons -- collective oscillations of free electron density -- in a charge-doped graphene is a very rapidly growing field \cite{Wunsch2006-318,Hwang2007-205418,Rana2008-91,Fei2011-4701,Velizhanin2011-085401,Koppens2011-3370,Bao2012-3677,Grigorenko2012-749,Fei2013-821,Luo2013-351,Low2014-1086}. Multiple potential applications of graphene plasmonics \cite{Grigorenko2012-749,Luo2013-351,Low2014-1086} are based or rely heavily upon the strong optical confinement and large density of states of graphene plasmons (GP), which is the consequence of the GP wavelength being typically much shorter than the photon wavelength at the same energy ($\lambda_{h\nu}/\lambda_{\rm GP}\sim20-100$)
\cite{Koppens2011-3370,Grigorenko2012-749}.

\begin{figure}
\centering
\includegraphics[width=3.2in]{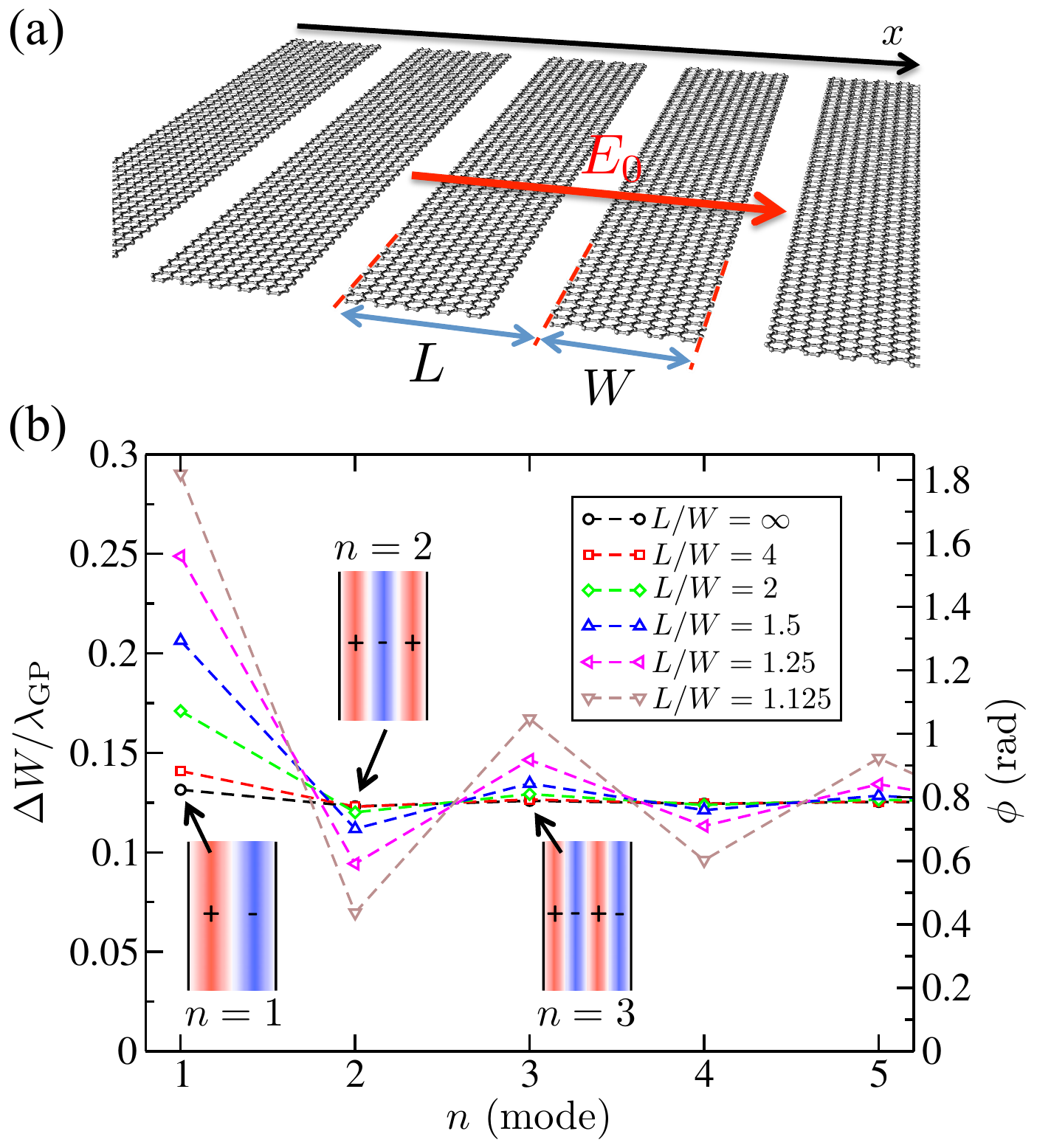}\caption{\label{fig:Schematic}(a) GNR array with $W$ and $L$ being the GNR
width and the width of the periodic unit, respectively. (b) Effective
extra width normalized to the plasmon wavelength (left vertical axis)
and the extra reflection phase (right vertical axis) vs the mode
index.}
\end{figure}

At these conditions, however, the simplest possible means of GP excitation, i.e., via photon absorption by a homogeneous graphene sheet, is not feasible since it is impossible to simultaneously conserve both energy and momentum. A lot of experimental and theoretical efforts have been devoted recently to the development of efficient optical and non-optical means to excite plasmons in graphene. Some of these efforts employed particles with dispersion relations sufficiently different from that of free photons, e.g., electrons \cite{Liu2008-201403,Liu2010-081406,Tegenkamp2011-012001,deAbajo2013-11409}, to be able to simultaneously conserve energy and momentum. Other efforts focused on breaking the continuous translational symmetry of the system, so that only energy has to be conserved. These include the formation of transient diffraction grating on the surface of graphene by launching acoustic waves \cite{Farhat2013-237404,Schiefele2013-237405}, as well as excitations of plasmons in near-field by a local defect like atomic force microscope (AFM) tip \cite{Fei2011-4701,Chen2012-77,Fei2012-82}, in-graphene impurities \cite{Muniz2010-081411}, or semiconductor quantum dot \cite{Koppens2011-3370,Velizhanin2011-085401}. The translational invariance can be broken not only by introduction of such {\it external} defects, but also by {\it nano-patterning} of graphene itself. Specifically, optical excitation of GPs in an array of graphene nanoribbons (GNR) has recently emerged
as one of the dominant experimental means to study GPs, Fig.~\ref{fig:Schematic}(a).
Size quantization of GPs in such nanoribbons gives rise to spatially localized
plasmon modes that readily couple to photons. Studies of GPs
using GNR arrays have already provided important insights into the
nature of plasmon damping in graphene and the efficiency of the plasmon coupling
to optical phonons in the surrounding material \cite{Ju2011-630,Yan2013-394,Brar2013-2541,Strait2013-241410,Abeysinghe2014-89932B}.

In order to use a GNR array to extract various properties of GPs (e.g., dispersion relation), an accurate theoretical description of plasmon resonances in (i) an isolated GNR and (ii) a GNR array is required. This has been addressed to some extent recently \cite{Christensen2012-431,Brar2013-2541,Strait2013-241410,deAbajo2014-135,Nikitin2014-041407R,Du2014-22689}, however no systematic study in this regard has been undertaken. In this work we  (i) systematically study the plasmonic response of periodic GNR arrays, and (ii)  provide a complete solution to the problem of size quantization of GPs in such systems. This solution can be directly used to analyze experimental results. 

Classically, a plasmon mode within a single GNR can be thought of
as a standing wave of charge ``sloshing'' perpendicular to the GNR
axis. The insets in Fig.~\ref{fig:Schematic}(b) show schematically the charge distribution for the three lowest-energy GP modes. Naively,
one would think that the boundary condition of vanishing electric
current at the GNR edges directly transforms into a reflection phase of $\pi$, resulting in a standard quantization condition
\begin{equation}
2Wk_{n}=2\pi n,~n=1,2,...\label{eq:scc}
\end{equation}
where $W$ is the GNR width and $k_{n}$ is the GP wavenumber
corresponding to the $n^{{\rm th}}$ mode. However, this is not entirely correct since just like any plasma oscillation, a GP consists of two coupled energy-carrying components: charge current and oscillating
electric field. The current-vanishing boundary condition does obviously apply only to the
current so that the electric field can effectively penetrate beyond the GNR edge resulting in a new quantization condition $2k_{n}W'_n=2\pi n$,
where
\begin{equation}
W'_n\equiv W+\Delta W_n>W \label{eq:Wp}
\end{equation}
is the $n$-dependent \emph{effective} GNR width. This condition can also be expressed as \cite{Brar2013-2541}
\begin{equation}
2k_{n}W+2\phi_n=2\pi n,\label{eq:quant}
\end{equation}
where $\phi_n=k_{n}\Delta W_n$ is an extra reflection phase accumulated
by a plasmon during the ``propagation'' outside a GNR. It turns out (see Sec.~\ref{sec:General}) that this phase is rather universal and depends {\em only} on the mode index $n$ and aspect ratio of a GNR array, $L/W$ (see Fig.~\ref{fig:Schematic}(a)). Therefore, evaluating $\phi_n$ for a few values of $n$ and $L/W$ (Fig.~\ref{fig:Schematic}(b) and Table~\ref{tab:phi_lambda}) provides a complete solution to the problem of GP size quantization in an arbitrary GNR array.

The paper is organized as follows. Section~\ref{sec:General} formalizes the problem of the polarization current in a GNR array in terms of an integro-differential equation. The spectral decomposition of the kernel of this equation provides an appealing geometric perspective onto the size quantization of plasmons in GNRs. A simple approximate theory of this size quantization is developed in Sec.~\ref{sec:analytics}. Section~\ref{sec:conclusion} concludes. 

\section{General Theory \& Spectral Decomposition}\label{sec:General}

From the onset we will limit ourselves to the situation where (i) graphene is assumed to be a purely two-dimensional ``zero-thickness'' material,  and the projection of the external electric field onto the graphene's plane is (ii) homogeneous,
$E_{0}(x,\omega)\equiv E_{0}(\omega)$, and (iii) polarized perpendicular to GNR axes, Fig.~\ref{fig:Schematic}(a).
At these conditions the problem becomes effectively one-dimensional and the polarization
current within a GNR array can be written as $j(x,\omega)=\sigma(x,\omega)\left[E_{0}(\omega)+E_{ind}(x,\omega)\right]$,
where $E_{ind}(x,\omega)$ is the induced electric field. The spatially-resolved
surface conductivity of a GNR array is denoted by $\sigma(x,\omega)$ \footnote{In general the surface conductivity is non-local, $\sigma({\bf x},{\bf x}';\omega)$. However, if the characteristic excitation (i.e., plasmon) wavelength is
much larger than the Fermi wavelength of graphene, then it can be
assumed that $\sigma({\bf x},{\bf x}';\omega)\approx\sigma({\bf x},\omega)\delta({\bf x}-{\bf x}')$.
We will assume this locality approximation henceforth.}.
Using the typically large ratio $\lambda_{h\nu}/\lambda_{\rm GP}\sim20-100$ one can neglect retardation effects and relate the induced electric field to the induced surface charge density of graphene, $\rho(x,\omega)$, via (in Gaussian units)
\begin{equation}
E_{ind}(x,\omega)=\int dx'\,\frac{2}{x-x'}\rho(x,\omega),
\end{equation}
where $\frac{2}{x-x'}$ is the electric field of a line charge with
unit linear density. The integration is assumed in the Cauchy principal
value sense. Using the expressions above and the continuity relation,
$-i\omega\rho(x,\omega)+\partial_{x}j(x,\omega)=0$, one can write
down a closed equation for the polarization current as
\begin{equation}
j(x)=\sigma(x)E_{0}-\frac{2i\sigma(x)}{\omega}\int dx'\,\frac{\partial_{x'}j(x')}{x-x'},\label{eq:IDE}
\end{equation}
where the explicit dependence on $\omega$ is omitted for brevity.

The obtained integro-differential equation can be straightforwardly
modified if the environment-induced dielectric screening is present,
which is the case when a GNR array is fabricated on top of some dielectric
substrate (e.g., ${\rm SiO_{2}}$). The effective dielectric constant
of environment, $\epsilon$, then enters the problem via a modified
electric field of a line charge, $\frac{2}{\epsilon(x-x')}$. This
modification is straightforwardly absorbed into $\sigma(x)$, which
is what is assumed in what follows. 

Equation~(\ref{eq:IDE}) can be solved numerically as a large system of
linear equations via discretization of $j(x)$ and $\sigma(x)$ on
a real-space or momentum-space grid, the latter based on the spatial Fourier transform of Eq.~(\ref{eq:IDE}). The real-space approach is most suitable
in the case of an isolated GNR (i.e., $L/W\rightarrow\infty$). The
momentum-space approach -- expansion of $j(x)$ and $\sigma(x)$ into plane
waves with periodic boundary conditions -- is ideal when $L$ is finite.
Indeed, using the momentum-space expansion with the period set to
$L$ one automatically obtains a solution for the \emph{infinite}
periodic GNR array so there is no need to solve a computationally
intensive problem of a very large but still finite number of GNRs within an array \cite{Strait2013-241410}.

\subsection{Spectral decomposition}

A more insightful and physically transparent approach to solving Eq.~(\ref{eq:IDE})
is to reformulate it as an eigenvalue problem. To this end we first
consider an integro-differential operator in the second r.h.s. term
of Eq.~(\ref{eq:IDE}). That this operator is not symmetric
complicates its spectral decomposition. However, by defining a new
unknown function as $y(x)=\sigma^{1/2}(x)E_{ind}(x)$, one obtains
a new equation
\begin{equation}
y(x)=\sigma^{1/2}(x)E_{0}-\frac{2i}{\omega}\int dx'\,\frac{\sigma^{1/2}(x)\partial_{x'}\left[\sigma^{1/2}(x')y(x')\right]}{x-x'},
\end{equation}
where the operator is now symmetric \footnote{This can be demonstrated by applying an integration by parts to an arbitrary off-diagonal matrix element of this operator}.
Further simplification can be obtained for a specific but very important
case where the spatial variation of the conductivity within the GNR
array can be expressed as
\begin{equation}
\sigma(x)=\sigma_{0}h(x),\label{eq:cond_step}
\end{equation}
where $h(x)=1$ when $x$ is within a GNR and $h(x)=0$ otherwise \footnote{A more general situation would be to have the conductivity changing continually from some finite value to zero at the GNR edge. We do not consider this situation in the present work.}.
Then, the integro-differential equation can be rewritten as
\begin{equation}
y(x)=\sigma^{1/2}(x)E_{0}-\frac{2i\pi\sigma_{0}}{\omega}(\hat{K}y)(x),\label{eq:IDEy}
\end{equation}
where the operator $\hat{K}$ is defined as
\begin{equation}
(\hat{K}y)(x)=\frac{1}{\pi}\int dx'\,\frac{h(x)\partial_{x'}\left[h(x')y(x')\right]}{x-x'}.\label{eq:Kop}
\end{equation}
This operator is real and symmetric so that it can be diagonalized
with all the eigenvalues being real and a set of eigenfunctions forming
a complete orthogonal basis. Therefore, in matrix bra-ket notation
this operator is decomposed as $\hat{K}=\sum_{n}k_{n}|y_n\rangle\langle y_n|$
so that Eq.~(\ref{eq:IDEy}) can be formally solved as
\begin{equation}
|y\rangle=\sum_{n}\frac{|y_n\rangle\langle y_n|}{\left(1+\frac{2\pi i\sigma_{0}}{\omega}k_{n}\right)}|\tilde{y}\rangle,
\end{equation}
where $\tilde{y}(x)=\sigma^{1/2}(x)E_{0}$. Restoring the original
real-space notation and multiplying both sides by $\sigma^{1/2}(x)$
we obtain the spatially-resolved polarization current as
\begin{equation}
j(x)=\sigma_{0}E_{0}\sum_{n}\frac{h(x)y_{n}(x)\int dx'\, h(x')y_{n}(x')}{\left(1+\frac{2\pi i\sigma_{0}}{\omega}k_{n}\right)}.\label{eq:jx}
\end{equation}
The homogeneous current, i.e., the one directly coupled the external homogeneous electric field, is given by $L_{tot}^{-1}\int dx\, j(x)$,
where $L_{tot}$ is the total length of GNR in $x$ direction. The effective homogeneous
conductivity of the system is then (recovering explicit frequency dependence)
\begin{equation}
\tilde{\sigma}(\omega)=\sigma_{0}(\omega)\sum_{n}\frac{\Lambda_{n}}{\left[1+\frac{2\pi i\sigma_{0}(\omega)}{\omega}k_{n}\right]},\label{eq:seff}
\end{equation}
where $\Lambda_{n}$ is given by
\begin{equation}
\Lambda_{n}=L_{tot}^{-1}\left[\int dx\, h(x)y_{n}(x)\right]^{2}.\label{eq:Lnc}
\end{equation}
As a function of $\omega$, $\tilde{\sigma}(\omega)$ can have resonances
when one of the denominators vanishes or nearly vanishes. The positions and the intensities
of such resonances are determined by $k_{n}$ and $\Lambda_{n}$,
respectively. The obtained universal spectral decomposition is similar in spirit to that obtained recently in Ref.~\cite{deAbajo2014-135}.

In the limit of continuous graphene, $h(x)\equiv 1$, one can show
that $y(x)=e^{ikx}$ with arbitrary real $k$ is an eigenfunction
of operator $\hat{K}$ with eigenvalue $|k|$. Therefore, eigenvalues
$k_{n}$ can be interpreted as effective wave numbers of a continuous
GP being size-quantized within a GNR array. As was discussed in Introduction, $k_{n}$ does not exactly match the ``naive'' size
quantization conditions since there is a finite phase accumulation,
$\phi_{n}=\pi n-k_{n}W>0$, that occurs when a GP ``propagates'' beyond
the GNR edge. The advantage of the solution of the problem given by
Eqs.~(\ref{eq:jx}) and (\ref{eq:seff}) is that operator $\hat{K}$
is purely geometric, i.e., it does depend only on the geometric configuration
of a GNR array via $h(x)$ but not on $\omega$ or $\sigma_0(\omega)$. Furthermore, a simple size rescaling of Eq.~(\ref{eq:Kop}) shows
that $\phi_{n}$ and $\Lambda_{n}$ are functions of only two parameters: $n$ and $L/W$, and not of $L$ or $W$ separately. Therefore, one can say that the plasmonic response of different GNR arrays with the same $L/W$ belong to the same {\em geometric universality class} since $\hat{K}$ -- the operator that encodes the geometry of a GNR array and determines the GP size quantization -- is exactly the same for them up to the size rescaling.

This geometric universality is a generalization of the previously introduced electrostatic scaling law \cite{Christensen2012-431}. The advantage of the former is that the diagonalization of operator $\hat{K}$, done only once for each value of $L/W$, gives not only the positions of resonance peaks but the full information on the frequency-resolved optical response of a GNR array via Eq.~(\ref{eq:seff}). This includes peak intensities as well as their widths and shapes.

Tabulating numerically evaluated $\phi_{n}$ and $\Lambda_{n}$ for a few first modes within a range of $L/W$ constitutes then, with the help of Eq.~(\ref{eq:seff}), the complete solution to the problem of size quantization of GPs in an arbitrary periodic GNR array. Table~\ref{tab:phi_lambda} gives the numerical values for the extra reflection phase and the resonance strength (in the form of $\lambda_{n}=n^{2}\frac{L}{W}\Lambda_{n}$) for the first three plasmon modes. The numerical diagonalization of operator $\hat{K}$ has been performed on a real-space grid for an isolated GNR ($L/W=\infty$) and using the plane wave basis set at finite $L/W$. The convergence with respect to the basis (or grid) size was thoroughly tested and $\sim10^3-10^4$ basis functions (grid points) were sufficient for the numerical convergence of all numerical results presented in this work. 
\begin{table}
\centering
\begin{tabular}{|c|c|c|c|c|c|c|}
\hline 
$L/W$ & $\phi_{1}$ & $\lambda_{1}$ & $\phi_{2}$ & $\lambda_{2}$ & $\phi_{3}$ & $\lambda_{3}$\tabularnewline
\hline 
\hline 
$\infty$ & 0.826 & 0.888 & 0.774 & 0 & 0.791 & 0.513\tabularnewline
\hline 
$4$ & 0.885 & 0.891 & 0.773 & 0 & 0.795 & 0.504\tabularnewline
\hline 
$2$ & 1.075 & 0.896 & 0.755 & 0 & 0.812 & 0.471\tabularnewline
\hline 
$1.5$ & 1.297 & 0.902 & 0.703 & 0 & 0.846 & 0.429\tabularnewline
\hline 
$1.25$ & 1.563 & 0.912 & 0.593 & 0 & 0.921 & 0.372\tabularnewline
\hline 
$1.125$ & 1.823 & 0.923 & 0.438 & 0 & 1.049 & 0.310\tabularnewline
\hline 
$1.05$ & 2.105 & 0.938 & 0.240 & 0 & 1.286 & 0.237\tabularnewline
\hline 
$1.01$ & 2.429 & 0.961 & 0.059 & 0 & 1.719 & 0.146\tabularnewline
\hline 
\end{tabular}
\caption{\label{tab:phi_lambda}The extra reflection phase ($\phi_{n}$) and
the resonance strength (in the form of $\lambda_{n}=n^{2}\Lambda_{n}\frac{L}{W}$)
tabulated for the first three plasmon modes within a range of aspect
ratios ($L/W$) of GNR arrays.}
\end{table}
The numerical results for the isolated GNR (the first line in the table) are consistent with those obtained very recently elsewhere \cite{Nikitin2014-041407R}.

To see if the extra reflection phase is significant it has to be compared
to the ``naive'' phase a GP accumulates when getting from one edge
of GNR to another, i.e., $\pi n$ for the $n^{{\rm th}}$ mode [see Eq.~(\ref{eq:scc})]. Naturally,
the correction is most significant for the first mode ($n=1$), for example
$\phi_{1}/\pi=0.263$ for an isolated GNR constitutes a significant
correction if a resonance frequency is used to draw some conclusions
on the plasmonic response of graphene, e.g., its dispersion relation.
Furthermore, one can notice that the difference between $\phi_{1}$
for an isolated GNR ($L/W\rightarrow\infty$) and for a GNR array
with $L/W=2$ is also non-negligible. At these conditions, GNRs in a very
typical experimental configuration \cite{Yan2013-394,Strait2013-241410,Brar2013-2541}
($L/W=2$) cannot be considered isolated and the interaction between
GNRs has to be accounted for by assuming $\phi_1=1.075$ and not $\phi_1=0.826$ as it was in the case of the isolated GNR. 

\section{Analytical Estimates}\label{sec:analytics}

\begin{figure}
\centering
\includegraphics[width=3.2in]{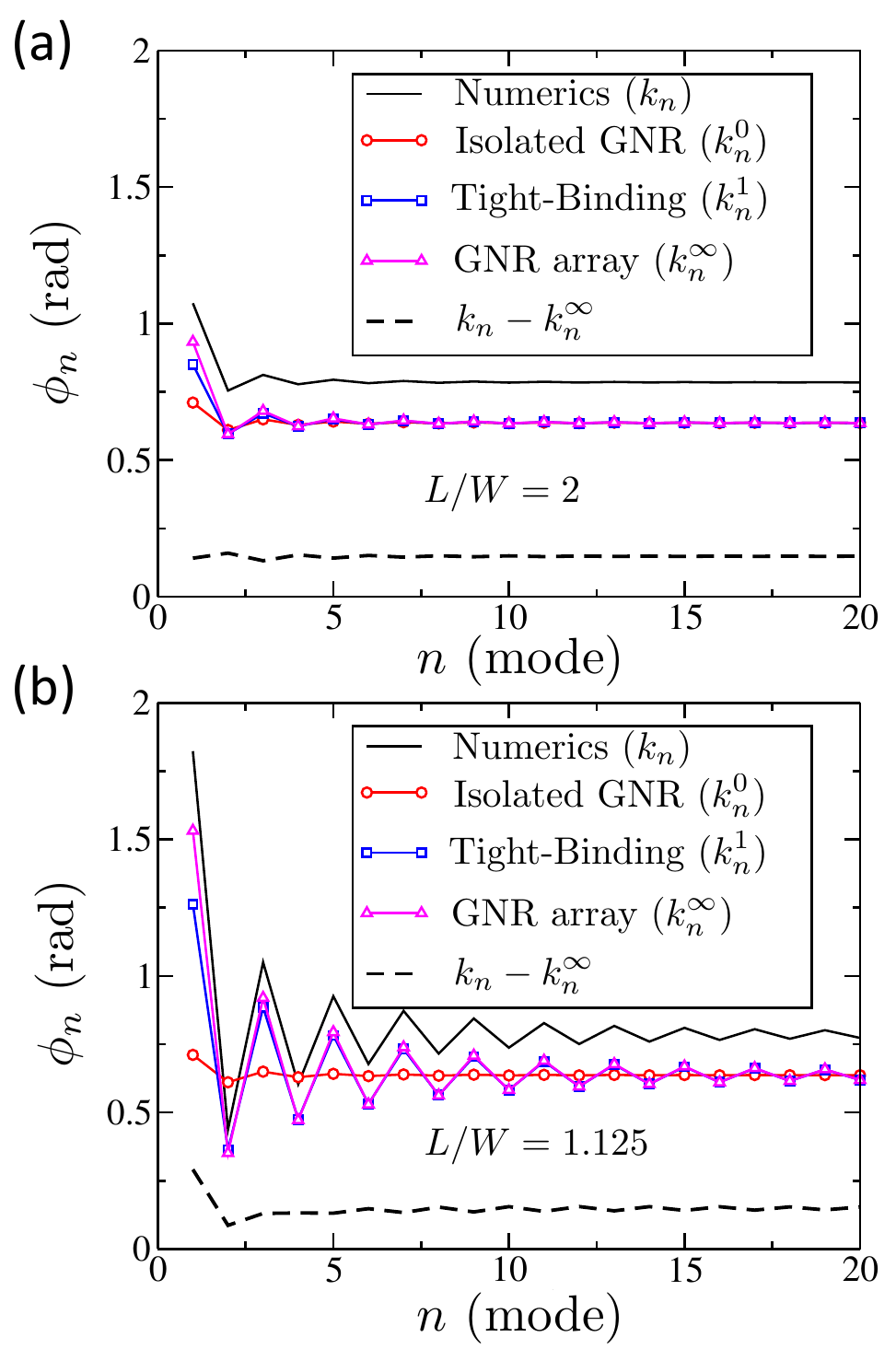}\caption{\label{fig:analytics} Comparison of numerically exact simulations
(solid black line) and the approximate analytical approach. (a) Typical
experimental configuration ($L/W=2)$. (b) ``Wide-nanoribbon-thin-slit''
configuration ($L/W=1.125$). In both panels, analytical results are
shown for: an isolated GNR ($k_{n}^{0}$, red circles), tight-binding
($k_{n}^{1}$, blue squares) and full GNR array ($k_{n}^{\infty}$, magenta
triangles). Black dashed line is the difference between the numerical
simulation results and $k_{n}^{\infty}$.}
\end{figure}

The problem of the plasmonic response of a GNR array has been solved
numerically in the previous section. Equation~(\ref{eq:seff}) parametrized
by data in Table~\ref{tab:phi_lambda} gives the frequency-resolved
effective conductivity of a GNR array. However, it would be great
to develop a simpler (i.e., analytical) theory to reproduce the trends
in the dependence of $\phi_{n}$ and $\Lambda_{n}$ on $n$ and $L/W$.
Such theory would be useful when simple ``quick-and-dirty'' estimates are needed and also if a deeper intuition on the physics of size quantization of GPs is required. To this end we assume a perturbative approach where as
a zeroth-order approximation we take the eigenfunctions of operator
$\hat{K}$ in Eq.~(\ref{eq:Kop}) to be simple (normalized) standing
waves, i.e.,
\begin{equation}
y_{n}^{(0)}(x)=\left(\frac{2L}{L_{tot}W}\right)^{1/2}\sum_{m}h_{m}(x)\sin\left[q_{n}(x+mL)\right],\label{eq:ssw}
\end{equation}
where $h_{m}(x)=1$ only when $x$ is within the $m^{{\rm th}}$ GNR,
and $q_{n}=\pi n/W$ comes from the ``naive'' size quantization. Then,
the first-order-corrected eigenvalues of $\hat{K}$ can be evaluated
as (in matrix notation) $k_{n}=\langle y_{n}|\hat{K}|y_{n}\rangle$ and
the explicit substitution of Eq.~(\ref{eq:ssw}) into this expression
yields
\begin{align}
k_{n}&=\frac{2Lq_{n}}{\pi L_{tot}W}\sum_{m,m'}\int_{0}^{W}dx\int_{0}^{W}dx'\,\frac{\sin(q_{n}x)\cos(q_{n}x')}{(x+mL)-(x'+m'L)}\nonumber\\
&=\frac{2}{\pi W}\sum_{m=-\infty}^{\infty}\int_{0}^{q_{n}W}dx\int_{0}^{q_{n}W}dx'\,\frac{\sin(x)\cos(x')}{x-x'+mq_{n}L}.
\end{align}
Performing substitution $u=x-x'$ and $v=(x+x')/2$ one obtains $k_{n}=\lim_{N\rightarrow\infty}k_{n}^{N}$,
where
\begin{equation}
k_{n}^{N}=\frac{2}{\pi W}\sum_{m=-N}^{N}\int_{0}^{q_{n}W}du\,\left(q_{n}W-u\right)\frac{\sin u}{u+mL}.
\end{equation}
The contribution $m=0$ to $k_{n}$ is easily evaluated as
\begin{equation}
k_{n}^{0}=\frac{2}{\pi W}\left[q_{n}W{\rm Si}(q_{n}W)+\cos q_{n}W-1\right],
\end{equation}
where ${\rm Si}(x)=\int^x_0 dt\,\frac{\sin t}{t}$ is the sine integral \cite{AbramowitzStegun1965}.
This is the final answer for an isolated GNR. If other GNRs are nearby
however, interaction with them has to be accounted for. To this end,
we first have to evaluate the following integral
\begin{align}
S(b,a)&\equiv\int_{0}^{a}du\,\frac{\sin u}{u+b}\nonumber\\
&={\rm Si}(a+b,b)\cos b-{\rm Ci}(|a+b|,|b|)\sin b,\label{eq:S}
\end{align}
where we define ${\rm Si}(a,b)\equiv{\rm Si}(a)-{\rm Si}(b)$ and ${\rm Ci}(a,b)\equiv{\rm Ci}(a)-{\rm Ci}(b)$. The cosine integral is given by ${\rm Ci}(x)=-\int^\infty_x dt\,\frac{\cos t}{t}$ \cite{AbramowitzStegun1965}. Equation~(\ref{eq:S}) is valid when (i) $a>0$, (ii)
$|b|>a$ (i.e., $b$ can be negative) or $b=0$. Then, $k_{n}^{N}$
defined above becomes
\begin{align}
k_{n}^{N}=\frac{2}{\pi W}\sum_{m=-N}^{N}&\left[q_{n}(W+mL)S(mq_{n}L;q_{n}W)\right.
\nonumber\\
&\left.+\cos q_{n}W-1\right].\label{eq:knn}
\end{align}
In Fig.~\ref{fig:analytics}, we compare these analytical results with
numerical simulations for a GNR array with $L/W=2$
[panel (a)] and $L/W=1.125$ [panel (b)]. 
Analytical calculations are
performed for the case where interaction between GNRs is turned off
($k_{n}^{0}$, red line), interaction only between nearest nanoribbons
is turned on ($k_{n}^{1}$, blue squares) and for the whole GNR array
where interaction between any two GNRs is allowed ($k_{n}^{\infty}$,
magenta triangles). As is seen, even in the case of GNRs separated
by a very thin slit [panel (b)] the tight-binding description is already
very close to the full analytical description ($k_{n}^{\infty}$). The latter
reproduces all the qualitative features of the exact numerical solution
such as convergence to a constant at large $n$, as well as the phase
and the amplitude of oscillations of $\phi_{n}$ versus $n$. The
largest disagreement between numerical and analytical results is a
systematic down shift of the latter. The difference between numerical
and analytical results, plotted by black dashed lines in both panels,
is seen to be essentially a constant $\Delta\phi\approx0.15$ independent
on $L/W$, except for very few lowest plasmon modes. Thus, in principle,
one can use the analytical expression shifted by this empirical correction
constant as a good approximation to exact numerical results.

The zeroth-order analytical estimate for the resonance strength reads as
\begin{equation}
\Lambda_{n}=L_{tot}^{-1}\left[\int dx\, y_{n}^{(0)}(x)\right]^{2},\label{eq:Ln}
\end{equation}
where $y_{n}^{(0)}(x)$ is given by Eq.~(\ref{eq:ssw}). The straightforward
evaluation of this integral produces
\begin{equation}
\Lambda_{n}=\frac{W}{L}\frac{8}{\pi^{2}n^{2}}\sin^{2}\left(\pi n/2\right).\label{eq:L0}
\end{equation}
As is seen, the resonance strength vanishes exactly for even modes ($n=2,4,...$). This is related to the symmetry of a GNR array with respect to the inversion $x\rightarrow -x$, which results in a definite parity state (even or odd) of each plasmon mode. Even modes have even parity of the charge density distribution, Fig.~\ref{fig:Schematic}(b), thus producing zero dipole moment and, therefore, vanishing resonance strength. This phenomenon is related to symmetry and thus true not only for the perturbative calculations but also for the exact numerical ones. In particular, this is the reason for vanishing resonance strength $\lambda_2$ in Table~\ref{tab:phi_lambda}.

\begin{figure}
\centering
\includegraphics[width=3.2in]{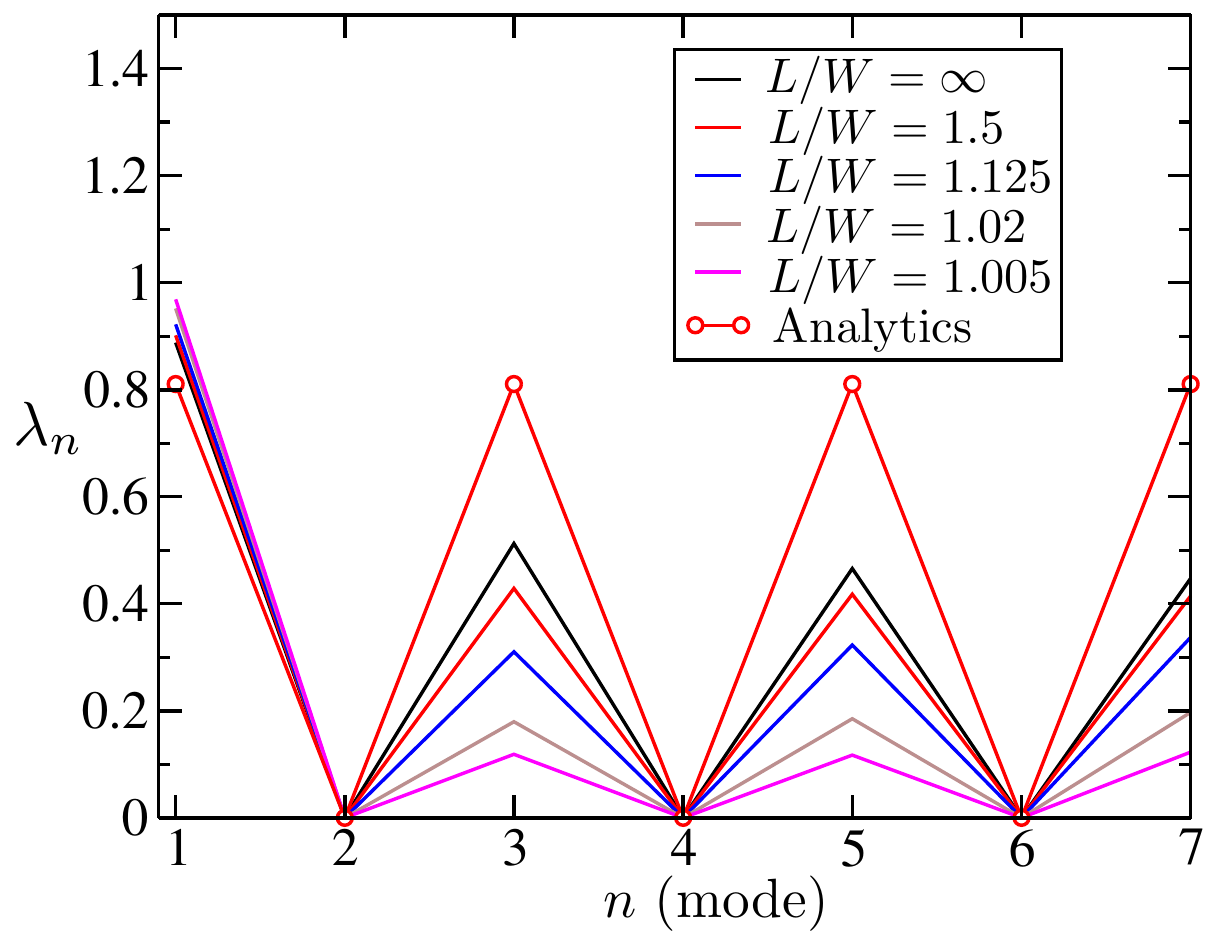}\caption{\label{fig:Lambda} Resonance strength (plotted as $\lambda_{n}\equiv\frac{Ln^{2}}{W}\Lambda_{n}$)
versus the index of the plasmon mode, $n$, for a range of aspect
ratios, $L/W$.}
\end{figure}

Fig.~\ref{fig:Lambda} shows the comparison of the analytical, Eq.~(\ref{eq:L0}), and numerical results for resonance strength plotted as $\lambda_{n}\equiv\frac{Ln^{2}}{W}\Lambda_{n}$.
It is seen that the analytics underestimates the resonance strength for all the modes except for the lowest one ($n=1$) by a factor of $\sim$2 at $L/W\rightarrow\infty$ (isolated GNR) and even more for
finite $L/W$. This observation can be rationalized by realizing that Eq.~(\ref{eq:L0}) is based on zeroth-order eigenfunctions of $\hat{K}$ so it does not account for coupling between GNRs. Decreasing $L/W$ results in stronger interaction between GNRs and thus leads to an increasing deviation of non-interacting analytical results from the numerical ones.

It is then rather counterintuitive that the analytics underestimates the resonance strength of the first plasmon mode {\em only slightly for all} $L/W$. This phenomenon can be explained using the sum
rule $\sum_{n=1}^{\infty}\Lambda_{n}=W/L$ that is applicable for
resonance strengths calculated from both the exact eigenfunctions
of $\hat{K}$ and the zeroth-order basis, Eqs.~(\ref{eq:ssw}) and
(\ref{eq:L0}). The derivation of this sum rule is given
in Appendix~\ref{app:sum_rule}. According to this sum rule,
analytically overestimating the resonance strengths for all the modes
with $n>1$ has to result in an underestimation of $\Lambda_1$ which is indeed the case. The reason why the analytical result
for $\Lambda_{1}$ is only slightly less than the numerical one
is that according to Eq.~(\ref{eq:L0}), $\Lambda_{n}$ decays rapidly
with $n$ so that most of the total resonance strength, $W/L$, has
to be concentrated in the very first resonance. Therefore, $\Lambda_{1}\approx W/L$ (i.e., $\lambda_{1}\approx1$) no matter which basis set of the two is used.

At $L/W\rightarrow1$ the resonance strengths of plasmon modes have
to decrease since $L/W=1$ corresponds to the case of homogeneous graphene where no plasmon can be excited by the homogeneous electric field assumed in this work. Resonance strengths at $n>1$ are indeed in agreement
with this expectation as is seen in Fig.~\ref{fig:Lambda}. However, the sum rule dictates that the total
resonance strength is conserved so that it becomes more and more concentrated
in the very first mode. Along with this, $\phi_{1}$ grows with $L/W\rightarrow1$
(see Table~\ref{tab:phi_lambda}) so that $k_{1}\rightarrow0$. These two observations lead to transformation of Eq.~(\ref{eq:seff}) into $\tilde{\sigma}(\omega)=\sigma_{0}(\omega)$
which is of course a quite expected result since the effective conductivity
of a uniform graphene has to reduce to its intrinsic conductivity, $\sigma_0(\omega)$.

\begin{figure}
\centering
\includegraphics[width=3.2in]{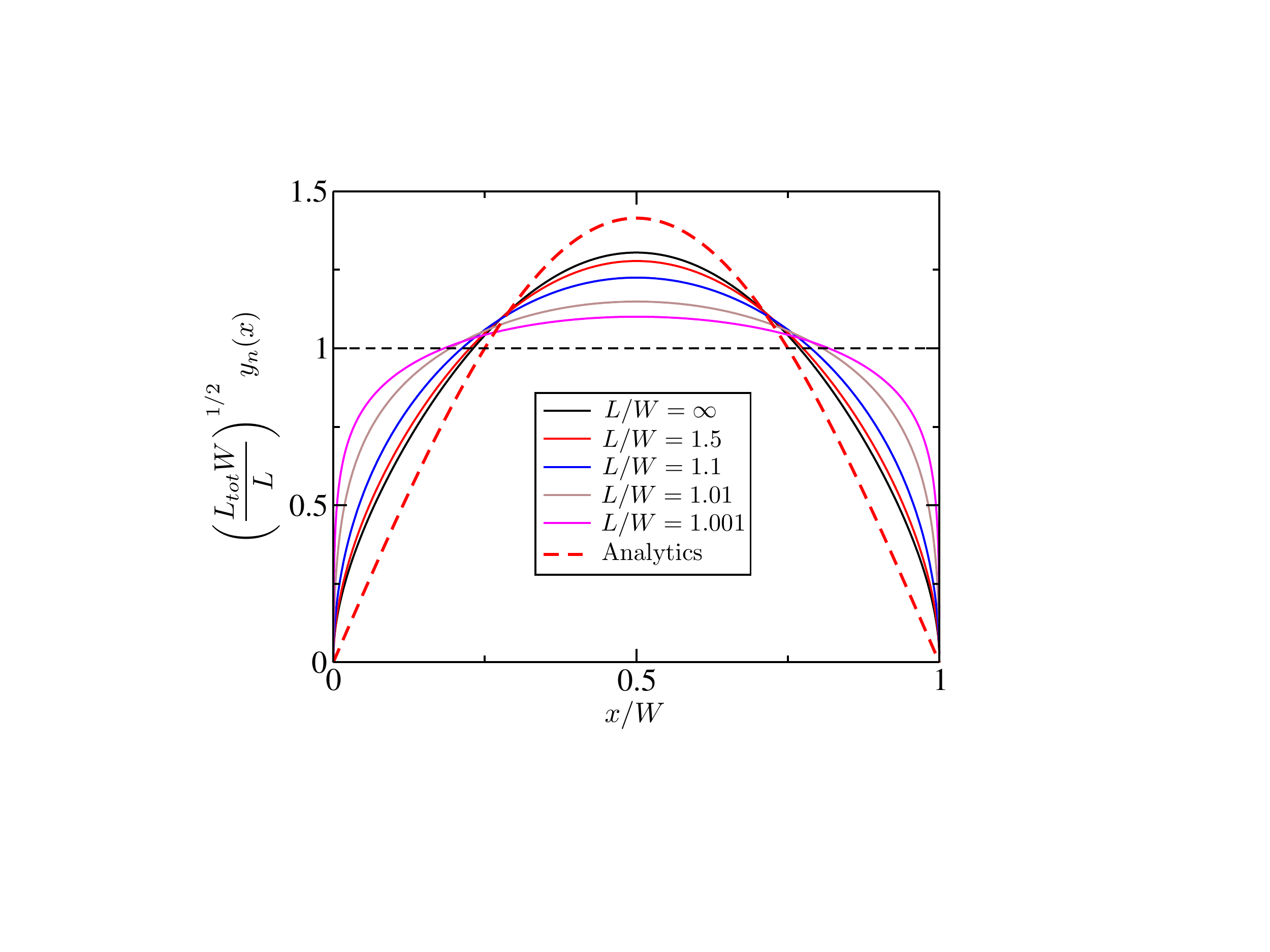}\caption{\label{fig:lmode} Eigenfunction of $\hat{K}$ corresponding to the lowest plasmon mode of a GNR array.}
\end{figure}

To further corroborate this, Fig.~\ref{fig:lmode} shows the profile of the lowest-mode eigenfunction for a range of $L/W$. 
For convenience, the eigenfunction is normalized with respect to a single GNR here and not to the entire GNR array, hence the prefactor $\left(\frac{L_{tot}W}{L}\right)^{1/2}$ for the vertical axis. The zeroth-order eigenfunction, $\sqrt{2}\sin(\pi x/W)$,
is shown by a dashed red line for comparison. One can see that this analytical dependence most closely resembles the eigenfunction of the isolated
GNR ($L/W=\infty$). Once the distance between GNRs decreases, the
numerically obtained eigenfunction deviates farther from the zeroth-order
one and approaches a constant value of 1 (shown by dashed black line
as a guide for the eye). Therefore, it is expected that in the limit of $L/W\rightarrow1$ the so normalized eigenfunction of the lowest plasmon mode will approach $1$ at any $x$ thus producing a homogeneous current consistent with $\tilde{\sigma}(\omega)=\sigma_{0}(\omega)$.

\section{Conclusion}\label{sec:conclusion}

In this paper, we analyze the problem of spatially confined plasmon modes in GNR arrays. We demonstrate that this problem can be decoupled into the problem of a plasmon in a uniform graphene sheet and the problem of size quantization. By focusing on the latter we demonstrate that such a size quantization is purely geometric in nature, i.e., the correct size quantization condition for plasmons in the GNR array is fully determined by the geometry of the array and nothing else. Further, we introduce the notion of the {\em geometric universality class} of GNR arrays where arrays with the same value of $L/W$ (other parameters are arbitrary) belong to the same class. The size quantization condition is universal within a class and can be obtained by a numerical (or approximate analytical) diagonalization of a certain integro-differential operator. We provide the results of accurate numerical diagonalization for the first three plasmon modes in Table~\ref{tab:phi_lambda}. The tabulated data can be directly used in analysis of experimental data on optical response of GNR arrays.

Finally, it is worthwhile to discuss the assumptions that were made in the very beginning of Sec.~\ref{sec:General}. From the perspective of dimensional analysis, since $L/W$ is the only dimensionless parameter of the problem, the only possible equation for the frequency of plasmon resonances is (neglecting the real component of graphene conductivity) \cite{Christensen2012-431}
\begin{equation}
\frac{{\rm Im}\left\{\sigma_0(\omega_n)\right\}}{\omega_n W}f(n,L/W)=1,\label{eq:DA}
\end{equation}
where $f(n,L/W)$ is a certain ``universal" function of $n$ and $L/W$. Introducing other parameters to the problem may lead to a more complex ``universal" function if extra dimensionless combinations can be constructed. For example, if the optical wavelength, $\lambda_{opt}$,  becomes comparable to $W$ in the frequency range of interest, then we are forced to introduce a new dimensionless parameter, $\lambda_{opt}/W$, as an argument of $f$. The same is true for, e.g., the Fermi wavelength in charge-doped graphene, $\lambda_F$, or the effective thickness of graphene sheet, $d$. Fortunately, one typically has $d,\lambda_F\ll W\ll\lambda_{opt}$ in realistic systems, so that graphene can be considered a truly two-dimensional system ($d\ll W$) with local conductivity ($\lambda_F\ll W$), and the homogeneous external electric field ($\lambda_{opt}\gg W$). At these conditions, dimensionless parameters $d/W$, $\lambda_F/W$ and $\lambda_{opt}/W$ are physically irrelevant bringing us back to Eq.~(\ref{eq:DA}).

Finally, it is worth mentioning that the phenomenon considered in this work, i.e., the penetration of the electric field beyond the GNR edges, is not the only possible source of the inequality $W'_n\neq W$ in Eq.~(\ref{eq:Wp}) when realistic experimental conditions are considered. The quality of GNR edges can also affect the effective GNR width. For example, specific parameters and experimental conditions of electron-beam lithography can result in over-exposed \cite{Yan2013-394} or under-exposed \cite{Abeysinghe2014-89932B} GNR edges, resulting in the width of {\it conducting} graphene within a GNR being lower or higher, respectively, than the apparent GNR width extracted from scanning electron microscopy (SEM) images. At first glance, this uncertainty in the GNR width renders the analysis present in this paper somewhat useless since one would have to notice a small change in $W'_n$ on top of $W$ that is not accurately known because of lithographic imperfections. However, we would like to emphasize that these two effects scale differently with such parameters as $W$, $L$, and $n$ (see Fig.~\ref{fig:Schematic}). Indeed, the lithography-induced variation of the GNR width, i.e., under- or over-exposure, is expected to be independent of these parameters. On the other hand, $\Delta W_n$ is seen in Fig.~\ref{fig:Schematic}(b) to be dependent on $n$ and $L/W$ so that these two effects can be experimentally distinguished and thus analyzed independently. In the present work, $W$ is always assumed to be the actual width of conducting graphene, Eq.~(\ref{eq:cond_step}), and not the apparent width seen in SEM images.     

We are thankful to Anatoly Efimov for multiple discussions and the help with the manuscript. This work was performed under the NNSA of the U.S. DOE at LANL under Contract No. DE-AC52-06NA25396.

\appendix

\section{Sum Rule for Resonance Strength\label{app:sum_rule}}

In this Appendix we will demonstrate that the total resonance strength
of all the plasmon modes in a GNR array equals to the areal fraction of
graphene in the array, i.e., $\sum_{n=1}^{\infty}\Lambda_{n}=W/L$,
where $\Lambda_{n}$ is defined by Eq.~(\ref{eq:Lnc}). To this end
it is most convenient to work in a discretized real-space representation
where a normalized eigenfunction $y_{n}(x)$ is represented by vector
$|y_{n}\rangle$ with components defined as $\left[|y_{n}\rangle\right]_{i}=\Delta x^{1/2}y_{n}(x_{i})$,
where $\Delta x=L_{tot}/M$ is the discretization step and $M$ is
the total number of real space discretization points. Here, the
prefactor of $\Delta x^{1/2}$ is required so that $|y_{n}\rangle$
is normalized in a vector sense, i.e., $\langle y_{n}|y_{n}\rangle=\sum_i y^2_n(x_i)=1$.
Within this discrete picture, Eq.~(\ref{eq:Lnc}) takes on the form
\begin{align}
\Lambda_{n}&=L_{tot}^{-1}\left(\sum_{i}h(x_{i})y_{n}(x_{i})\Delta x\right)^{2}\nonumber\\
&=L_{tot}^{-1}\left(\Delta x^{1/2}\langle h|y_{n}\rangle\right)^{2},
\end{align}
where $\left[|h\rangle\right]_{i}=h(x_{i})$. The summation over all
the resonance strengths then becomes
\begin{equation}
\sum_{n}\Lambda_{n}=L_{tot}^{-1}\Delta x\sum_{n}\langle y_{n}|h\rangle\langle h|y_{n}\rangle.\label{eq:sumLn}
\end{equation}
In this expression, the summation over the complete orthogonal basis
$\{y_{n}\}$ is equivalent to evaluation of the trace of matrix $\hat{h}=|h\rangle\langle h|$.
A diagonal elements of this matrix, $h_{ii}$, equals to $1$ if $x_{i}$
corresponds to a position within a GNR, and $0$ otherwise. Therefore,
the trace of $\hat{h}$ is proportional to the areal fraction of graphene
in the GNR array. More specifically, $\sum_{n}\langle y_{n}|h\rangle\langle h|y_{n}\rangle={\rm Tr}\hat{h}=M\frac{W}{L}$.
Substituting this result into Eq.~(\ref{eq:sumLn}) one obtains
\begin{equation}
\sum_{n}\Lambda_{n}=W/L.\label{eq:sumLnf}
\end{equation}
This result does not depend on the basis $y_{n}$ as long as it is
complete. For example, the basis does not have to consist of exact
eigenfunctions of operator $\hat{K}$ for Eq.~(\ref{eq:sumLnf})
to be true. Furthermore, if the basis is not complete in the entire
space but complete in the space defined by equation $h(x)=1$, it
still produces Eq.~(\ref{eq:sumLnf}). Therefore, the summation of
zeroth-order resonance strengths, Eq.~(\ref{eq:L0}), still produces
$W/L$ since the zeroth-order basis, Eq.~(\ref{eq:ssw}) is complete
on GNRs. This can also be shown by direct summation.


\begin{thebibliography}{33}%
\makeatletter
\providecommand \@ifxundefined [1]{%
 \@ifx{#1\undefined}
}%
\providecommand \@ifnum [1]{%
 \ifnum #1\expandafter \@firstoftwo
 \else \expandafter \@secondoftwo
 \fi
}%
\providecommand \@ifx [1]{%
 \ifx #1\expandafter \@firstoftwo
 \else \expandafter \@secondoftwo
 \fi
}%
\providecommand \natexlab [1]{#1}%
\providecommand \enquote  [1]{``#1''}%
\providecommand \bibnamefont  [1]{#1}%
\providecommand \bibfnamefont [1]{#1}%
\providecommand \citenamefont [1]{#1}%
\providecommand \href@noop [0]{\@secondoftwo}%
\providecommand \href [0]{\begingroup \@sanitize@url \@href}%
\providecommand \@href[1]{\@@startlink{#1}\@@href}%
\providecommand \@@href[1]{\endgroup#1\@@endlink}%
\providecommand \@sanitize@url [0]{\catcode `\\12\catcode `\$12\catcode
  `\&12\catcode `\#12\catcode `\^12\catcode `\_12\catcode `\%12\relax}%
\providecommand \@@startlink[1]{}%
\providecommand \@@endlink[0]{}%
\providecommand \url  [0]{\begingroup\@sanitize@url \@url }%
\providecommand \@url [1]{\endgroup\@href {#1}{\urlprefix }}%
\providecommand \urlprefix  [0]{URL }%
\providecommand \Eprint [0]{\href }%
\providecommand \doibase [0]{http://dx.doi.org/}%
\providecommand \selectlanguage [0]{\@gobble}%
\providecommand \bibinfo  [0]{\@secondoftwo}%
\providecommand \bibfield  [0]{\@secondoftwo}%
\providecommand \translation [1]{[#1]}%
\providecommand \BibitemOpen [0]{}%
\providecommand \bibitemStop [0]{}%
\providecommand \bibitemNoStop [0]{.\EOS\space}%
\providecommand \EOS [0]{\spacefactor3000\relax}%
\providecommand \BibitemShut  [1]{\csname bibitem#1\endcsname}%
\let\auto@bib@innerbib\@empty
\bibitem [{\citenamefont {Wunsch}\ \emph {et~al.}(2006)\citenamefont {Wunsch},
  \citenamefont {Stauber}, \citenamefont {Sols},\ and\ \citenamefont
  {Guinea}}]{Wunsch2006-318}%
  \BibitemOpen
  \bibfield  {author} {\bibinfo {author} {\bibfnamefont {B.}~\bibnamefont
  {Wunsch}}, \bibinfo {author} {\bibfnamefont {T.}~\bibnamefont {Stauber}},
  \bibinfo {author} {\bibfnamefont {F.}~\bibnamefont {Sols}}, \ and\ \bibinfo
  {author} {\bibfnamefont {F.}~\bibnamefont {Guinea}},\ }\href@noop {}
  {\bibfield  {journal} {\bibinfo  {journal} {New. J. Phys.}\ }\textbf
  {\bibinfo {volume} {8}},\ \bibinfo {pages} {318} (\bibinfo {year}
  {2006})}\BibitemShut {NoStop}%
\bibitem [{\citenamefont {Hwang}\ and\ \citenamefont
  {Das~Sarma}(2007)}]{Hwang2007-205418}%
  \BibitemOpen
  \bibfield  {author} {\bibinfo {author} {\bibfnamefont {E.~H.}\ \bibnamefont
  {Hwang}}\ and\ \bibinfo {author} {\bibfnamefont {S.}~\bibnamefont
  {Das~Sarma}},\ }\href@noop {} {\bibfield  {journal} {\bibinfo  {journal}
  {Phys. Rev. B}\ }\textbf {\bibinfo {volume} {75}},\ \bibinfo {pages} {205418}
  (\bibinfo {year} {2007})}\BibitemShut {NoStop}%
\bibitem [{\citenamefont {Rana}(2008)}]{Rana2008-91}%
  \BibitemOpen
  \bibfield  {author} {\bibinfo {author} {\bibfnamefont {F.}~\bibnamefont
  {Rana}},\ }\href@noop {} {\bibfield  {journal} {\bibinfo  {journal} {IEEE
  Trans. Nanotech.}\ }\textbf {\bibinfo {volume} {7}},\ \bibinfo {pages} {91}
  (\bibinfo {year} {2008})}\BibitemShut {NoStop}%
\bibitem [{\citenamefont {Fei}\ \emph {et~al.}(2011)\citenamefont {Fei},
  \citenamefont {Andreev}, \citenamefont {Bao}, \citenamefont {Zhang},
  \citenamefont {McLeod}, \citenamefont {Wang}, \citenamefont {Stewart},
  \citenamefont {Zhao}, \citenamefont {Dominguez}, \citenamefont {Thiemens},
  \citenamefont {Fogler}, \citenamefont {Tauber}, \citenamefont {Castro-Neto},
  \citenamefont {Lau}, \citenamefont {Keilmann},\ and\ \citenamefont
  {Basov}}]{Fei2011-4701}%
  \BibitemOpen
  \bibfield  {author} {\bibinfo {author} {\bibfnamefont {Z.}~\bibnamefont
  {Fei}}, \bibinfo {author} {\bibfnamefont {G.~O.}\ \bibnamefont {Andreev}},
  \bibinfo {author} {\bibfnamefont {W.}~\bibnamefont {Bao}}, \bibinfo {author}
  {\bibfnamefont {L.~M.}\ \bibnamefont {Zhang}}, \bibinfo {author}
  {\bibfnamefont {A.~S.}\ \bibnamefont {McLeod}}, \bibinfo {author}
  {\bibfnamefont {C.}~\bibnamefont {Wang}}, \bibinfo {author} {\bibfnamefont
  {M.~K.}\ \bibnamefont {Stewart}}, \bibinfo {author} {\bibfnamefont
  {Z.}~\bibnamefont {Zhao}}, \bibinfo {author} {\bibfnamefont {G.}~\bibnamefont
  {Dominguez}}, \bibinfo {author} {\bibfnamefont {M.}~\bibnamefont {Thiemens}},
  \bibinfo {author} {\bibfnamefont {M.~M.}\ \bibnamefont {Fogler}}, \bibinfo
  {author} {\bibfnamefont {M.~J.}\ \bibnamefont {Tauber}}, \bibinfo {author}
  {\bibfnamefont {A.~H.}\ \bibnamefont {Castro-Neto}}, \bibinfo {author}
  {\bibfnamefont {C.~N.}\ \bibnamefont {Lau}}, \bibinfo {author} {\bibfnamefont
  {F.}~\bibnamefont {Keilmann}}, \ and\ \bibinfo {author} {\bibfnamefont
  {D.~N.}\ \bibnamefont {Basov}},\ }\href@noop {} {\bibfield  {journal}
  {\bibinfo  {journal} {Nano letters}\ }\textbf {\bibinfo {volume} {11}},\
  \bibinfo {pages} {4701} (\bibinfo {year} {2011})}\BibitemShut {NoStop}%
\bibitem [{\citenamefont {Velizhanin}\ and\ \citenamefont
  {Efimov}(2011)}]{Velizhanin2011-085401}%
  \BibitemOpen
  \bibfield  {author} {\bibinfo {author} {\bibfnamefont {K.~A.}\ \bibnamefont
  {Velizhanin}}\ and\ \bibinfo {author} {\bibfnamefont {A.}~\bibnamefont
  {Efimov}},\ }\href@noop {} {\bibfield  {journal} {\bibinfo  {journal} {Phys.
  Rev. B}\ }\textbf {\bibinfo {volume} {84}},\ \bibinfo {pages} {085401}
  (\bibinfo {year} {2011})}\BibitemShut {NoStop}%
\bibitem [{\citenamefont {Koppens}\ \emph {et~al.}(2011)\citenamefont
  {Koppens}, \citenamefont {Chang},\ and\ \citenamefont {Garcia~de
  Abajo}}]{Koppens2011-3370}%
  \BibitemOpen
  \bibfield  {author} {\bibinfo {author} {\bibfnamefont {F.~H.}\ \bibnamefont
  {Koppens}}, \bibinfo {author} {\bibfnamefont {D.~E.}\ \bibnamefont {Chang}},
  \ and\ \bibinfo {author} {\bibfnamefont {F.~J.}\ \bibnamefont {Garcia~de
  Abajo}},\ }\href@noop {} {\bibfield  {journal} {\bibinfo  {journal} {Nano
  Lett.}\ }\textbf {\bibinfo {volume} {11}},\ \bibinfo {pages} {3370} (\bibinfo
  {year} {2011})}\BibitemShut {NoStop}%
\bibitem [{\citenamefont {Bao}\ and\ \citenamefont {Loh}(2012)}]{Bao2012-3677}%
  \BibitemOpen
  \bibfield  {author} {\bibinfo {author} {\bibfnamefont {Q.}~\bibnamefont
  {Bao}}\ and\ \bibinfo {author} {\bibfnamefont {K.~P.}\ \bibnamefont {Loh}},\
  }\href@noop {} {\bibfield  {journal} {\bibinfo  {journal} {ACS Nano}\
  }\textbf {\bibinfo {volume} {6}},\ \bibinfo {pages} {3677} (\bibinfo {year}
  {2012})}\BibitemShut {NoStop}%
\bibitem [{\citenamefont {Grigorenko}\ \emph {et~al.}(2012)\citenamefont
  {Grigorenko}, \citenamefont {Polini},\ and\ \citenamefont
  {Novoselov}}]{Grigorenko2012-749}%
  \BibitemOpen
  \bibfield  {author} {\bibinfo {author} {\bibfnamefont {A.~N.}\ \bibnamefont
  {Grigorenko}}, \bibinfo {author} {\bibfnamefont {M.}~\bibnamefont {Polini}},
  \ and\ \bibinfo {author} {\bibfnamefont {K.~S.}\ \bibnamefont {Novoselov}},\
  }\href@noop {} {\bibfield  {journal} {\bibinfo  {journal} {Nature Phot.}\
  }\textbf {\bibinfo {volume} {6}},\ \bibinfo {pages} {749} (\bibinfo {year}
  {2012})}\BibitemShut {NoStop}%
\bibitem [{\citenamefont {Z.~Fei}\ \emph {et~al.}(2013)\citenamefont {Z.~Fei},
  \citenamefont {Rodin}, \citenamefont {Gannett}, \citenamefont {Dai},
  \citenamefont {Regan}, \citenamefont {Wagner}, \citenamefont {Liu},
  \citenamefont {McLeod}, \citenamefont {Dominguez}, \citenamefont {Thiemens},
  \citenamefont {Castro~Neto}, \citenamefont {Keilmann}, \citenamefont {Zettl},
  \citenamefont {Hillenbrand}, \citenamefont {Fogler},\ and\ \citenamefont
  {Basov}}]{Fei2013-821}%
  \BibitemOpen
  \bibfield  {author} {\bibinfo {author} {\bibfnamefont {Z.}~\bibnamefont
  {Z.~Fei}}, \bibinfo {author} {\bibfnamefont {A.~S.}\ \bibnamefont {Rodin}},
  \bibinfo {author} {\bibfnamefont {W.}~\bibnamefont {Gannett}}, \bibinfo
  {author} {\bibfnamefont {S.}~\bibnamefont {Dai}}, \bibinfo {author}
  {\bibfnamefont {W.}~\bibnamefont {Regan}}, \bibinfo {author} {\bibfnamefont
  {M.}~\bibnamefont {Wagner}}, \bibinfo {author} {\bibfnamefont {M.~K.}\
  \bibnamefont {Liu}}, \bibinfo {author} {\bibfnamefont {A.~S.}\ \bibnamefont
  {McLeod}}, \bibinfo {author} {\bibfnamefont {G.}~\bibnamefont {Dominguez}},
  \bibinfo {author} {\bibfnamefont {M.}~\bibnamefont {Thiemens}}, \bibinfo
  {author} {\bibfnamefont {A.~H.}\ \bibnamefont {Castro~Neto}}, \bibinfo
  {author} {\bibfnamefont {F.}~\bibnamefont {Keilmann}}, \bibinfo {author}
  {\bibfnamefont {A.}~\bibnamefont {Zettl}}, \bibinfo {author} {\bibfnamefont
  {R.}~\bibnamefont {Hillenbrand}}, \bibinfo {author} {\bibfnamefont {M.~M.}\
  \bibnamefont {Fogler}}, \ and\ \bibinfo {author} {\bibfnamefont {D.~N.}\
  \bibnamefont {Basov}},\ }\href@noop {} {\bibfield  {journal} {\bibinfo
  {journal} {Nature Nanotech.}\ }\textbf {\bibinfo {volume} {8}},\ \bibinfo
  {pages} {821} (\bibinfo {year} {2013})}\BibitemShut {NoStop}%
\bibitem [{\citenamefont {Luo}\ \emph {et~al.}(2013)\citenamefont {Luo},
  \citenamefont {Qiu}, \citenamefont {Lu},\ and\ \citenamefont
  {Ni}}]{Luo2013-351}%
  \BibitemOpen
  \bibfield  {author} {\bibinfo {author} {\bibfnamefont {X.}~\bibnamefont
  {Luo}}, \bibinfo {author} {\bibfnamefont {T.}~\bibnamefont {Qiu}}, \bibinfo
  {author} {\bibfnamefont {W.}~\bibnamefont {Lu}}, \ and\ \bibinfo {author}
  {\bibfnamefont {Z.}~\bibnamefont {Ni}},\ }\href@noop {} {\bibfield  {journal}
  {\bibinfo  {journal} {Mater. Sci. Eng. R-Rep.}\ }\textbf {\bibinfo {volume}
  {74}},\ \bibinfo {pages} {351} (\bibinfo {year} {2013})}\BibitemShut
  {NoStop}%
\bibitem [{\citenamefont {Low}\ and\ \citenamefont
  {Avouris}(2014)}]{Low2014-1086}%
  \BibitemOpen
  \bibfield  {author} {\bibinfo {author} {\bibfnamefont {T.}~\bibnamefont
  {Low}}\ and\ \bibinfo {author} {\bibfnamefont {P.}~\bibnamefont {Avouris}},\
  }\href@noop {} {\bibfield  {journal} {\bibinfo  {journal} {ACS Nano}\
  }\textbf {\bibinfo {volume} {8}},\ \bibinfo {pages} {1086} (\bibinfo {year}
  {2014})}\BibitemShut {NoStop}%
\bibitem [{\citenamefont {Liu}\ \emph {et~al.}(2008)\citenamefont {Liu},
  \citenamefont {Willis}, \citenamefont {Emtsev},\ and\ \citenamefont
  {Seyller}}]{Liu2008-201403}%
  \BibitemOpen
  \bibfield  {author} {\bibinfo {author} {\bibfnamefont {Y.}~\bibnamefont
  {Liu}}, \bibinfo {author} {\bibfnamefont {R.~F.}\ \bibnamefont {Willis}},
  \bibinfo {author} {\bibfnamefont {K.~V.}\ \bibnamefont {Emtsev}}, \ and\
  \bibinfo {author} {\bibfnamefont {T.}~\bibnamefont {Seyller}},\ }\href@noop
  {} {\bibfield  {journal} {\bibinfo  {journal} {Phys. Rev. B}\ }\textbf
  {\bibinfo {volume} {78}},\ \bibinfo {pages} {201403(R)} (\bibinfo {year}
  {2008})}\BibitemShut {NoStop}%
\bibitem [{\citenamefont {Liu}\ and\ \citenamefont
  {Willis}(2010)}]{Liu2010-081406}%
  \BibitemOpen
  \bibfield  {author} {\bibinfo {author} {\bibfnamefont {Y.}~\bibnamefont
  {Liu}}\ and\ \bibinfo {author} {\bibfnamefont {R.~F.}\ \bibnamefont
  {Willis}},\ }\href@noop {} {\bibfield  {journal} {\bibinfo  {journal} {Phys.
  Rev. B}\ }\textbf {\bibinfo {volume} {81}},\ \bibinfo {pages} {081406(R)}
  (\bibinfo {year} {2010})}\BibitemShut {NoStop}%
\bibitem [{\citenamefont {Tegenkamp}\ \emph {et~al.}(2011)\citenamefont
  {Tegenkamp}, \citenamefont {Pfnur}, \citenamefont {Langer}, \citenamefont
  {Baringhaus},\ and\ \citenamefont {Schumacher}}]{Tegenkamp2011-012001}%
  \BibitemOpen
  \bibfield  {author} {\bibinfo {author} {\bibfnamefont {C.}~\bibnamefont
  {Tegenkamp}}, \bibinfo {author} {\bibfnamefont {H.}~\bibnamefont {Pfnur}},
  \bibinfo {author} {\bibfnamefont {T.}~\bibnamefont {Langer}}, \bibinfo
  {author} {\bibfnamefont {J.}~\bibnamefont {Baringhaus}}, \ and\ \bibinfo
  {author} {\bibfnamefont {H.~W.}\ \bibnamefont {Schumacher}},\ }\href@noop {}
  {\bibfield  {journal} {\bibinfo  {journal} {J. Phys.: Cond. Mat.}\ }\textbf
  {\bibinfo {volume} {23}},\ \bibinfo {pages} {012001} (\bibinfo {year}
  {2011})}\BibitemShut {NoStop}%
\bibitem [{\citenamefont {Garcia~de Abajo}(2013)}]{deAbajo2013-11409}%
  \BibitemOpen
  \bibfield  {author} {\bibinfo {author} {\bibfnamefont {F.~J.}\ \bibnamefont
  {Garcia~de Abajo}},\ }\href@noop {} {\bibfield  {journal} {\bibinfo
  {journal} {ACS Nano}\ }\textbf {\bibinfo {volume} {7}},\ \bibinfo {pages}
  {11409} (\bibinfo {year} {2013})}\BibitemShut {NoStop}%
\bibitem [{\citenamefont {Farhat}\ \emph {et~al.}(2013)\citenamefont {Farhat},
  \citenamefont {Guenneau},\ and\ \citenamefont {Bagci}}]{Farhat2013-237404}%
  \BibitemOpen
  \bibfield  {author} {\bibinfo {author} {\bibfnamefont {M.}~\bibnamefont
  {Farhat}}, \bibinfo {author} {\bibfnamefont {S.}~\bibnamefont {Guenneau}}, \
  and\ \bibinfo {author} {\bibfnamefont {H.}~\bibnamefont {Bagci}},\
  }\href@noop {} {\bibfield  {journal} {\bibinfo  {journal} {Phys. Rev. Lett.}\
  }\textbf {\bibinfo {volume} {111}},\ \bibinfo {pages} {237404} (\bibinfo
  {year} {2013})}\BibitemShut {NoStop}%
\bibitem [{\citenamefont {Schiefele}\ \emph {et~al.}(2013)\citenamefont
  {Schiefele}, \citenamefont {Pedros}, \citenamefont {Sols}, \citenamefont
  {Calle},\ and\ \citenamefont {Guinea}}]{Schiefele2013-237405}%
  \BibitemOpen
  \bibfield  {author} {\bibinfo {author} {\bibfnamefont {J.}~\bibnamefont
  {Schiefele}}, \bibinfo {author} {\bibfnamefont {J.}~\bibnamefont {Pedros}},
  \bibinfo {author} {\bibfnamefont {F.}~\bibnamefont {Sols}}, \bibinfo {author}
  {\bibfnamefont {F.}~\bibnamefont {Calle}}, \ and\ \bibinfo {author}
  {\bibfnamefont {F.}~\bibnamefont {Guinea}},\ }\href@noop {} {\bibfield
  {journal} {\bibinfo  {journal} {Phys. Rev. Lett.}\ }\textbf {\bibinfo
  {volume} {111}},\ \bibinfo {pages} {237405} (\bibinfo {year}
  {2013})}\BibitemShut {NoStop}%
\bibitem [{\citenamefont {Chen}\ \emph {et~al.}(2012)\citenamefont {Chen},
  \citenamefont {Badioli}, \citenamefont {Alonso-Gonzalez}, \citenamefont
  {Thongrattanasiri}, \citenamefont {Huth}, \citenamefont {Osmond},
  \citenamefont {Spasenovic}, \citenamefont {Centeno}, \citenamefont
  {Pesquera}, \citenamefont {Godignon}, \citenamefont {Elorza}, \citenamefont
  {Camara}, \citenamefont {Garcia~de Abajo}, \citenamefont {Hillenbrand},\ and\
  \citenamefont {Koppens}}]{Chen2012-77}%
  \BibitemOpen
  \bibfield  {author} {\bibinfo {author} {\bibfnamefont {J.}~\bibnamefont
  {Chen}}, \bibinfo {author} {\bibfnamefont {M.}~\bibnamefont {Badioli}},
  \bibinfo {author} {\bibfnamefont {P.}~\bibnamefont {Alonso-Gonzalez}},
  \bibinfo {author} {\bibfnamefont {S.}~\bibnamefont {Thongrattanasiri}},
  \bibinfo {author} {\bibfnamefont {F.}~\bibnamefont {Huth}}, \bibinfo {author}
  {\bibfnamefont {J.}~\bibnamefont {Osmond}}, \bibinfo {author} {\bibfnamefont
  {M.}~\bibnamefont {Spasenovic}}, \bibinfo {author} {\bibfnamefont
  {A.}~\bibnamefont {Centeno}}, \bibinfo {author} {\bibfnamefont
  {A.}~\bibnamefont {Pesquera}}, \bibinfo {author} {\bibfnamefont
  {P.}~\bibnamefont {Godignon}}, \bibinfo {author} {\bibfnamefont {A.~Z.}\
  \bibnamefont {Elorza}}, \bibinfo {author} {\bibfnamefont {N.}~\bibnamefont
  {Camara}}, \bibinfo {author} {\bibfnamefont {F.~J.}\ \bibnamefont {Garcia~de
  Abajo}}, \bibinfo {author} {\bibfnamefont {R.}~\bibnamefont {Hillenbrand}}, \
  and\ \bibinfo {author} {\bibfnamefont {F.~H.}\ \bibnamefont {Koppens}},\
  }\href@noop {} {\bibfield  {journal} {\bibinfo  {journal} {Nature}\ }\textbf
  {\bibinfo {volume} {487}},\ \bibinfo {pages} {77} (\bibinfo {year}
  {2012})}\BibitemShut {NoStop}%
\bibitem [{\citenamefont {Fei}\ \emph {et~al.}(2012)\citenamefont {Fei},
  \citenamefont {Rodin}, \citenamefont {Andreev}, \citenamefont {Bao},
  \citenamefont {McLeod}, \citenamefont {Wagner}, \citenamefont {Zhang},
  \citenamefont {Zhao}, \citenamefont {Thiemens}, \citenamefont {Dominguez},
  \citenamefont {Fogler}, \citenamefont {Castro~Neto}, \citenamefont {Lau},
  \citenamefont {Keilmann},\ and\ \citenamefont {Basov}}]{Fei2012-82}%
  \BibitemOpen
  \bibfield  {author} {\bibinfo {author} {\bibfnamefont {Z.}~\bibnamefont
  {Fei}}, \bibinfo {author} {\bibfnamefont {A.~S.}\ \bibnamefont {Rodin}},
  \bibinfo {author} {\bibfnamefont {G.~O.}\ \bibnamefont {Andreev}}, \bibinfo
  {author} {\bibfnamefont {W.}~\bibnamefont {Bao}}, \bibinfo {author}
  {\bibfnamefont {A.~S.}\ \bibnamefont {McLeod}}, \bibinfo {author}
  {\bibfnamefont {M.}~\bibnamefont {Wagner}}, \bibinfo {author} {\bibfnamefont
  {L.~M.}\ \bibnamefont {Zhang}}, \bibinfo {author} {\bibfnamefont
  {Z.}~\bibnamefont {Zhao}}, \bibinfo {author} {\bibfnamefont {M.}~\bibnamefont
  {Thiemens}}, \bibinfo {author} {\bibfnamefont {G.}~\bibnamefont {Dominguez}},
  \bibinfo {author} {\bibfnamefont {M.~M.}\ \bibnamefont {Fogler}}, \bibinfo
  {author} {\bibfnamefont {A.~H.}\ \bibnamefont {Castro~Neto}}, \bibinfo
  {author} {\bibfnamefont {C.~N.}\ \bibnamefont {Lau}}, \bibinfo {author}
  {\bibfnamefont {F.}~\bibnamefont {Keilmann}}, \ and\ \bibinfo {author}
  {\bibfnamefont {D.~N.}\ \bibnamefont {Basov}},\ }\href@noop {} {\bibfield
  {journal} {\bibinfo  {journal} {Nature}\ }\textbf {\bibinfo {volume} {487}},\
  \bibinfo {pages} {82} (\bibinfo {year} {2012})}\BibitemShut {NoStop}%
\bibitem [{\citenamefont {Muniz}\ \emph {et~al.}(2010)\citenamefont {Muniz},
  \citenamefont {Dahal}, \citenamefont {Balatsky},\ and\ \citenamefont
  {Haas}}]{Muniz2010-081411}%
  \BibitemOpen
  \bibfield  {author} {\bibinfo {author} {\bibfnamefont {R.~A.}\ \bibnamefont
  {Muniz}}, \bibinfo {author} {\bibfnamefont {H.~P.}\ \bibnamefont {Dahal}},
  \bibinfo {author} {\bibfnamefont {A.~V.}\ \bibnamefont {Balatsky}}, \ and\
  \bibinfo {author} {\bibfnamefont {S.}~\bibnamefont {Haas}},\ }\href@noop {}
  {\bibfield  {journal} {\bibinfo  {journal} {Phys. Rev. B}\ }\textbf {\bibinfo
  {volume} {82}},\ \bibinfo {pages} {081411(R)} (\bibinfo {year}
  {2010})}\BibitemShut {NoStop}%
\bibitem [{\citenamefont {Ju}\ \emph {et~al.}(2011)\citenamefont {Ju},
  \citenamefont {Geng}, \citenamefont {Horng}, \citenamefont {Girit},
  \citenamefont {Martin}, \citenamefont {Hao}, \citenamefont {Bechtel},
  \citenamefont {Liang}, \citenamefont {Zettl}, \citenamefont {Shen},\ and\
  \citenamefont {Wang}}]{Ju2011-630}%
  \BibitemOpen
  \bibfield  {author} {\bibinfo {author} {\bibfnamefont {L.}~\bibnamefont
  {Ju}}, \bibinfo {author} {\bibfnamefont {B.}~\bibnamefont {Geng}}, \bibinfo
  {author} {\bibfnamefont {J.}~\bibnamefont {Horng}}, \bibinfo {author}
  {\bibfnamefont {C.}~\bibnamefont {Girit}}, \bibinfo {author} {\bibfnamefont
  {M.}~\bibnamefont {Martin}}, \bibinfo {author} {\bibfnamefont
  {Z.}~\bibnamefont {Hao}}, \bibinfo {author} {\bibfnamefont {H.~A.}\
  \bibnamefont {Bechtel}}, \bibinfo {author} {\bibfnamefont {X.}~\bibnamefont
  {Liang}}, \bibinfo {author} {\bibfnamefont {A.}~\bibnamefont {Zettl}},
  \bibinfo {author} {\bibfnamefont {Y.~R.}\ \bibnamefont {Shen}}, \ and\
  \bibinfo {author} {\bibfnamefont {F.}~\bibnamefont {Wang}},\ }\href@noop {}
  {\bibfield  {journal} {\bibinfo  {journal} {Nature Nanotech.}\ }\textbf
  {\bibinfo {volume} {6}},\ \bibinfo {pages} {630} (\bibinfo {year}
  {2011})}\BibitemShut {NoStop}%
\bibitem [{\citenamefont {Yan}\ \emph {et~al.}(2013)\citenamefont {Yan},
  \citenamefont {Low}, \citenamefont {Zhu}, \citenamefont {Wu}, \citenamefont
  {Freitag}, \citenamefont {Li}, \citenamefont {Guinea}, \citenamefont
  {Avouris},\ and\ \citenamefont {Xia}}]{Yan2013-394}%
  \BibitemOpen
  \bibfield  {author} {\bibinfo {author} {\bibfnamefont {H.}~\bibnamefont
  {Yan}}, \bibinfo {author} {\bibfnamefont {T.}~\bibnamefont {Low}}, \bibinfo
  {author} {\bibfnamefont {W.}~\bibnamefont {Zhu}}, \bibinfo {author}
  {\bibfnamefont {Y.}~\bibnamefont {Wu}}, \bibinfo {author} {\bibfnamefont
  {M.}~\bibnamefont {Freitag}}, \bibinfo {author} {\bibfnamefont
  {X.}~\bibnamefont {Li}}, \bibinfo {author} {\bibfnamefont {F.}~\bibnamefont
  {Guinea}}, \bibinfo {author} {\bibfnamefont {P.}~\bibnamefont {Avouris}}, \
  and\ \bibinfo {author} {\bibfnamefont {F.}~\bibnamefont {Xia}},\ }\href@noop
  {} {\bibfield  {journal} {\bibinfo  {journal} {Nature Phot.}\ }\textbf
  {\bibinfo {volume} {7}},\ \bibinfo {pages} {394} (\bibinfo {year}
  {2013})}\BibitemShut {NoStop}%
\bibitem [{\citenamefont {Brar}\ \emph {et~al.}(2013)\citenamefont {Brar},
  \citenamefont {Jang}, \citenamefont {Sherrott}, \citenamefont {Lopez},\ and\
  \citenamefont {Atwater}}]{Brar2013-2541}%
  \BibitemOpen
  \bibfield  {author} {\bibinfo {author} {\bibfnamefont {V.~W.}\ \bibnamefont
  {Brar}}, \bibinfo {author} {\bibfnamefont {M.~S.}\ \bibnamefont {Jang}},
  \bibinfo {author} {\bibfnamefont {M.}~\bibnamefont {Sherrott}}, \bibinfo
  {author} {\bibfnamefont {J.~J.}\ \bibnamefont {Lopez}}, \ and\ \bibinfo
  {author} {\bibfnamefont {H.~A.}\ \bibnamefont {Atwater}},\ }\href@noop {}
  {\bibfield  {journal} {\bibinfo  {journal} {Nano Lett.}\ }\textbf {\bibinfo
  {volume} {13}},\ \bibinfo {pages} {2541} (\bibinfo {year}
  {2013})}\BibitemShut {NoStop}%
\bibitem [{\citenamefont {Strait}\ \emph {et~al.}(2013)\citenamefont {Strait},
  \citenamefont {Nene}, \citenamefont {Chan}, \citenamefont {Manolatou},
  \citenamefont {Tiwari}, \citenamefont {Rana}, \citenamefont {Kevek},\ and\
  \citenamefont {McEuen}}]{Strait2013-241410}%
  \BibitemOpen
  \bibfield  {author} {\bibinfo {author} {\bibfnamefont {J.~H.}\ \bibnamefont
  {Strait}}, \bibinfo {author} {\bibfnamefont {P.}~\bibnamefont {Nene}},
  \bibinfo {author} {\bibfnamefont {W.-M.}\ \bibnamefont {Chan}}, \bibinfo
  {author} {\bibfnamefont {C.}~\bibnamefont {Manolatou}}, \bibinfo {author}
  {\bibfnamefont {S.}~\bibnamefont {Tiwari}}, \bibinfo {author} {\bibfnamefont
  {F.}~\bibnamefont {Rana}}, \bibinfo {author} {\bibfnamefont {J.~W.}\
  \bibnamefont {Kevek}}, \ and\ \bibinfo {author} {\bibfnamefont {P.~L.}\
  \bibnamefont {McEuen}},\ }\href@noop {} {\bibfield  {journal} {\bibinfo
  {journal} {Phys. Rev. B}\ }\textbf {\bibinfo {volume} {87}},\ \bibinfo
  {pages} {241410(R)} (\bibinfo {year} {2013})}\BibitemShut {NoStop}%
\bibitem [{\citenamefont {Abeysinghe}\ \emph {et~al.}(2013)\citenamefont
  {Abeysinghe}, \citenamefont {Myers}, \citenamefont {Nader~Esfahani},
  \citenamefont {Hendrickson}, \citenamefont {Cleary}, \citenamefont {Walker},
  \citenamefont {Chen}, \citenamefont {Chen},\ and\ \citenamefont
  {Mou}}]{Abeysinghe2014-89932B}%
  \BibitemOpen
  \bibfield  {author} {\bibinfo {author} {\bibfnamefont {D.~C.}\ \bibnamefont
  {Abeysinghe}}, \bibinfo {author} {\bibfnamefont {J.}~\bibnamefont {Myers}},
  \bibinfo {author} {\bibfnamefont {N.}~\bibnamefont {Nader~Esfahani}},
  \bibinfo {author} {\bibfnamefont {J.~R.}\ \bibnamefont {Hendrickson}},
  \bibinfo {author} {\bibfnamefont {J.~W.}\ \bibnamefont {Cleary}}, \bibinfo
  {author} {\bibfnamefont {D.~E.}\ \bibnamefont {Walker}}, \bibinfo {author}
  {\bibfnamefont {K.-H.}\ \bibnamefont {Chen}}, \bibinfo {author}
  {\bibfnamefont {L.-C.}\ \bibnamefont {Chen}}, \ and\ \bibinfo {author}
  {\bibfnamefont {S.}~\bibnamefont {Mou}},\ }\href@noop {} {\bibfield
  {journal} {\bibinfo  {journal} {Proc. of SPIE}\ }\textbf {\bibinfo {volume}
  {8993}},\ \bibinfo {pages} {89932B} (\bibinfo {year} {2013})}\BibitemShut
  {NoStop}%
\bibitem [{\citenamefont {Christensen}\ \emph {et~al.}(2012)\citenamefont
  {Christensen}, \citenamefont {Manjavacas}, \citenamefont {Thongrattanasiri},
  \citenamefont {Koppens},\ and\ \citenamefont {Garcia~de
  Abajo}}]{Christensen2012-431}%
  \BibitemOpen
  \bibfield  {author} {\bibinfo {author} {\bibfnamefont {J.}~\bibnamefont
  {Christensen}}, \bibinfo {author} {\bibfnamefont {A.}~\bibnamefont
  {Manjavacas}}, \bibinfo {author} {\bibfnamefont {S.}~\bibnamefont
  {Thongrattanasiri}}, \bibinfo {author} {\bibfnamefont {F.~H.~L.}\
  \bibnamefont {Koppens}}, \ and\ \bibinfo {author} {\bibfnamefont {F.~J.}\
  \bibnamefont {Garcia~de Abajo}},\ }\href@noop {} {\bibfield  {journal}
  {\bibinfo  {journal} {ACS Nano}\ }\textbf {\bibinfo {volume} {6}},\ \bibinfo
  {pages} {431} (\bibinfo {year} {2012})}\BibitemShut {NoStop}%
\bibitem [{\citenamefont {Garcia~de Abajo}(2014)}]{deAbajo2014-135}%
  \BibitemOpen
  \bibfield  {author} {\bibinfo {author} {\bibfnamefont {F.~J.}\ \bibnamefont
  {Garcia~de Abajo}},\ }\href@noop {} {\bibfield  {journal} {\bibinfo
  {journal} {ACS Photonics}\ }\textbf {\bibinfo {volume} {1}},\ \bibinfo
  {pages} {135} (\bibinfo {year} {2014})}\BibitemShut {NoStop}%
\bibitem [{\citenamefont {Nikitin}\ \emph {et~al.}(2014)\citenamefont
  {Nikitin}, \citenamefont {Low},\ and\ \citenamefont
  {Martin-Moreno}}]{Nikitin2014-041407R}%
  \BibitemOpen
  \bibfield  {author} {\bibinfo {author} {\bibfnamefont {A.~Y.}\ \bibnamefont
  {Nikitin}}, \bibinfo {author} {\bibfnamefont {T.}~\bibnamefont {Low}}, \ and\
  \bibinfo {author} {\bibfnamefont {L.}~\bibnamefont {Martin-Moreno}},\
  }\href@noop {} {\bibfield  {journal} {\bibinfo  {journal} {Phys. Rev. B}\
  }\textbf {\bibinfo {volume} {90}},\ \bibinfo {pages} {041407(R)} (\bibinfo
  {year} {2014})}\BibitemShut {NoStop}%
\bibitem [{\citenamefont {Du}\ \emph {et~al.}(2014)\citenamefont {Du},
  \citenamefont {Tang},\ and\ \citenamefont {Yuan}}]{Du2014-22689}%
  \BibitemOpen
  \bibfield  {author} {\bibinfo {author} {\bibfnamefont {L.}~\bibnamefont
  {Du}}, \bibinfo {author} {\bibfnamefont {D.}~\bibnamefont {Tang}}, \ and\
  \bibinfo {author} {\bibfnamefont {X.}~\bibnamefont {Yuan}},\ }\href@noop {}
  {\bibfield  {journal} {\bibinfo  {journal} {Opt. Express}\ }\textbf {\bibinfo
  {volume} {22}},\ \bibinfo {pages} {22689} (\bibinfo {year}
  {2014})}\BibitemShut {NoStop}%
\bibitem [{Note1()}]{Note1}%
  \BibitemOpen
  \bibinfo {note} {In general the surface conductivity is non-local, $\sigma
  ({\protect \bf x},{\protect \bf x}';\omega )$. However, if the characteristic
  excitation (i.e., plasmon) wavelength is much larger than the Fermi
  wavelength of graphene, then it can be assumed that $\sigma ({\protect \bf
  x},{\protect \bf x}';\omega )\approx \sigma ({\protect \bf x},\omega )\delta
  ({\protect \bf x}-{\protect \bf x}')$. We will assume this locality
  approximation henceforth.}\BibitemShut {Stop}%
\bibitem [{Note2()}]{Note2}%
  \BibitemOpen
  \bibinfo {note} {This can be demonstrated by applying an integration by parts
  to an arbitrary off-diagonal matrix element of this operator}\BibitemShut
  {NoStop}%
\bibitem [{Note3()}]{Note3}%
  \BibitemOpen
  \bibinfo {note} {A more general situation would be to have the conductivity
  changing continually from some finite value to zero at the GNR edge. We do
  not consider this situation in the present work.}\BibitemShut {Stop}%
\bibitem [{\citenamefont {Abramowitz}\ and\ \citenamefont
  {Stegun}(1965)}]{AbramowitzStegun1965}%
  \BibitemOpen
  \bibfield  {author} {\bibinfo {author} {\bibfnamefont {M.}~\bibnamefont
  {Abramowitz}}\ and\ \bibinfo {author} {\bibfnamefont {I.~A.}\ \bibnamefont
  {Stegun}},\ }\href@noop {} {\emph {\bibinfo {title} {Handbook of Mathematical
  Functions}}}\ (\bibinfo  {publisher} {Dover},\ \bibinfo {address} {New
  York},\ \bibinfo {year} {1965})\BibitemShut {NoStop}%
\end{thebibliography}

%

\end{document}